\documentclass[aip,reprint]{revtex4-1}

\usepackage{dcolumn}
\usepackage{bm}
\usepackage{mathrsfs}
\usepackage{amsmath}
\usepackage{braket}
\usepackage{graphicx,color}
\usepackage{epsfig}
\usepackage{natbib}
\usepackage[dvipsnames]{xcolor}

\begin{document}
\title{Revisiting the magnetic structure of Holmium at high pressure: a neutron diffraction study}
\author{M. Pardo-Sainz}
\affiliation{Instituto de Nanociencia y Materiales de Arag\'on (CSIC -- Universidad de Zaragoza) and depto. de F\'isica de la Materia Condensada, C/Pedro Cerbuna 12, 50009 Zaragoza, Spain.}
\affiliation{Graduate School of Science, Osaka Metropolitan University, 1-1 Gakuencho, Sakai, Osaka 599-8531 Japan.}
\author{F. Cova}
\affiliation{Instituto de Nanociencia y Materiales de Arag\'on (CSIC -- Universidad de Zaragoza) and depto. de F\'isica de la Materia Condensada, C/Pedro Cerbuna 12, 50009 Zaragoza, Spain.}
%
\author{J. A. Rodr\'iguez-Velamaz\'an}
\affiliation{Institut Laue Langevin, 71 avenue des Martyrs, CS 20156, Grenoble, Cedex 9 38042, France.}
\author{I.~Puente-Orench}
\affiliation{Instituto de Nanociencia y Materiales de Arag\'on (CSIC -- Universidad de Zaragoza) and depto. de F\'isica de la Materia Condensada, C/Pedro Cerbuna 12, 50009 Zaragoza, Spain.}
%
\author{Y. Kousaka}
\affiliation{Department of Physics and Electronics, Osaka Metropolitan University, 1-1 Gakuencho, Sakai, Osaka 599-8531 Japan.}
\author{M. Mito}
\email{mitoh@mns.kyutech.ac.jp}
\affiliation{Graduate School of Engineering, Kyushu Institute of Technology, Kitakyushu 804-8550, Japan.}
\author{J. Campo}
\email{javier.campo@csic.es}
\affiliation{Instituto de Nanociencia y Materiales de Arag\'on (CSIC -- Universidad de Zaragoza) and depto. de F\'isica de la Materia Condensada, C/Pedro Cerbuna 12, 50009 Zaragoza, Spain.}
\date{\today}
\begin{abstract}
\noindent
Low-temperature neutron diffraction experiments at $P= 8$ GPa have been conducted to investigate the magnetic structures of metallic Holmium at high pressures by employing a long $d$-spacing high-flux diffractometer and a Paris-Edinburgh press cell inside a cryostat. We find that at $P=8$ GPa and $T=5$ K, no nuclear symmetry change is observed, keeping therefore the hexagonal closed packed ($hcp$) symmetry at high pressure.  Our neutron diffraction data confirm that the ferromagnetic state does not exist. The magnetic structure corresponding to the helimagnetic order, which survives down to 5 K, is fully described by the magnetic superspace group formalism. These results are consistent with those previously published using magnetization experiments. 
\end{abstract}
\maketitle
\section{\label{sec:Intro}Introduction}
Ferromagnetic metals have played an important role in condensed matter physics from the viewpoint of its magnetism originated from itinerant electrons~\cite{Chikazumi_book}. 
In the 3$d$-electron ferromagnetic transition metals based on Fe, Co, and Ni, the mechanism responsible for their ferromagnetism can be understood within the Stoner model ~\cite{Stoner_1939}. 

On the other hand, the ferromagnetism in the 4$f$-electron lanthanide metals, such as Gd, Tb, Dy, Ho, Er, and Tm, is explained by the Ruderman-Kasuya-Kittle-Yosida (RKKY) interaction between localized moments of the $f$-orbital electrons mediated by the conduction electrons~\cite{RKKY_RK, RKKY_K, RKKY_Y}. 
The spatially damped oscillation of the conduction electron spin polarization is responsible of the competition between the ferromagnetic (FM) and antiferromagnetic (AFM) interactions, which often results in an incommensurate helimagnetic structure (HM).   

At low temperatures, the subtle lattice contractions modify the RKKY interaction ($J_{\rm RKKY}$) and the HM state is destabilized giving place to a FM ground state. Hereafter, the magnetic transition temperatures between the FM and HM states and between the HM and the paramagnetic (PM) states are denoted as $T_{\rm C}$ and $T_{\rm N}$, respectively.

All 4$f$-lanthanide FM metals have an hexagonal closed packed, \textit{hcp}, structure with stacking unit ABA at ambient pressure (AP), and exhibit the structural series of transformations in the sequence \textit{hcp} (ABA)  $\to$ Sm-type (ABABCBCACA) $\to$ double-\textit{hcp} (\textit{dhcp}) (ABACA) $\to$ \textit{fcc} (ABCA) $\to$ trigonal under increasing pressure.~\cite{Gd_structural_transition_1988, Gd-Ho_structural_transition_1992}

The variation of the magnetic properties with the structural transformations in 4$f$-lanthanide metals has been studied theoretically \cite{HUGHES2007} and reported experimentally by magnetic characterization~\cite{McWhan_1965, Iwamoto_2003, Jackson_2005, Gd_Tb_Dy_Ho_DAC, Gd_Tb_Dy_Ho_DAC_2021}, electrical resistivity~\cite{THOMAS2012, Gd_R_2014, Tb_R_neutron_pressure_Vohra, Dy_R_pressure_Vohra, Lim_2015_Gd_Dy, Lim_2015_Tb, Lim_2017_Gd_Tb_Dy}, neutron diffraction~\cite{Tb_Ho_neutron_pressure_1968, Tb_neutron_pressure_1992, Tb_R_neutron_pressure_Vohra, THOMAS2012, Dy_neutron_pressure, Dy_neutron_pressure_2022, Ho_neutron_pressure}, X-ray diffraction~\cite{CUNNINGHAM2006}, and M\"ossbauer spectroscopy~\cite{Dy_Mossbauer_XANES}. In particular, neutron diffraction experiments have been successfully employed to study the magnetic phases of Ho metal at high pressures and variable temperatures, as we briefly summarize in the next paragraph. 

The first neutron scattering experiment in this metal was performed at AP by Koehler \textit{et al.} \cite{KOEHLER1966}. In it, they reported that the Ho magnetic moments form a basal plane helix below $T_{\rm N}=133$ K, and a conical configuration was developed below $T_{\rm C}=20$ K, with a net magnetic moment parallel to the $c$-axis, in agreement with later neutron experiments \cite{PECHAN1984,SIMPSON1995}. In year 1968, Umebayashi \textit{et al.}~\cite{Tb_Ho_neutron_pressure_1968} studied Tb and Ho at pressures below 1 GPa and temperatures above 80 K, where the pressure dependences of $T_{\rm N}$ and the helical turn angle were measured. It was found that the HM ordering occurs at lower temperatures with increasing pressure. In 1988, Achiwa \textit{et al.}~\cite{ACHIWA1988} studied the Ho metal up to 2.1 GPa in the temperature range 10 K to $T_{\rm N}$. The helical pitch angle evolution with temperature was found in agreement with Ref.~\cite{Tb_Ho_neutron_pressure_1968} for $P=0.6$ GPa, while for higher pressures the values of the angle increased, and a lock-in value appeared below 20 K. 

Recently, the new available extreme conditions neutron diffractometers have facilitated the study of the high pressure region of Holmium. In 2012, Thomas \textit{et al.} ~\cite{THOMAS2012} performed neutron diffraction experiments at maximum pressures of 6.6 GPa at 89, 110 and 300 K. They established the incommensurate nature of the HM phase and determined the decrease of $T_{\rm N}$ from approximately 122~K at AP at a rate of -4.9 K/GPa up to a pressure of 9 GPa, above which the PM to HM transition vanishes, in agreement with Ref.~\cite{Tb_Ho_neutron_pressure_1968,ACHIWA1988}. In 2020, Perreault \textit{et al.}~\cite{Ho_neutron_pressure} performed neutron diffraction experiments at maximum pressures of 20 GPa and minimum temperatures of 10 K. They observed two magnetic transitions below 10 GPa: one to an incommensurate HM phase and another to a conical FM phase. For pressures above 10 GPa in the Sm-type phase, and above 19 GPa in the $dhcp$ phase, the appearance of a magnetic peak at 3 \AA\, and the increase of the intensity of some nuclear peaks were assigned to the presence of a FM ordering below 30 K. 

\begin{figure}[htb!]
\begin{center}
\includegraphics[width=0.85\columnwidth]{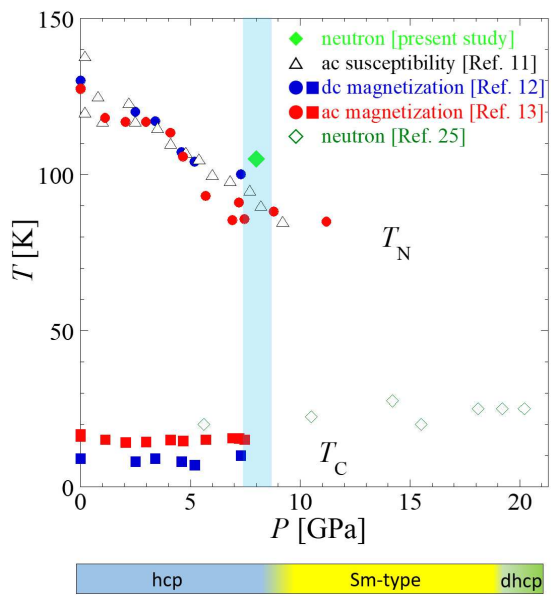}
\end{center}
\caption{\label{Figure-0} (Colour online) $P$ dependence of $T_{\rm N}$ and $T_{\rm C}$ for Ho~\cite{Jackson_2005,Gd_Tb_Dy_Ho_DAC,Gd_Tb_Dy_Ho_DAC_2021}.  The green point at $P=8$ GPa and $T_{\rm N}=105$~K has been determined in this study.  Blue and red symbols correspond to the data obtained from our previous magnetic characterization studies using an SQUID magnetometer, while the black empty triangles and green empty diamonds correspond to the previous a.c.~susceptibility and neutron diffraction experiments, respectively. The size of the error bars, in temperature and pressure, is smaller than the size of the symbols employed.  The color bar at the bottom part of the figure indicates the different phase transformations happening as the pressure increases at room temperature according to ref.~\cite{samudrala_structural_2013}. The present study focuses on the light-blue shaded region.}
\end{figure}

The $P$ dependence of $T_{\rm N}$ and $T_{\rm C}$, evaluated with magnetic susceptibility~\cite{Jackson_2005} (black-empty symbols), a.c and d.c~magnetization~\cite{Gd_Tb_Dy_Ho_DAC, Gd_Tb_Dy_Ho_DAC_2021} (blue and red filled symbols), and neutron diffraction~\cite{Ho_neutron_pressure} (green-empty symbols) is showed for Ho in the magnetic phase diagram in  Fig.~\ref{Figure-0}.

In our recent a.c.~magnetization measurements, using a superconducting quantum interference device (SQUID) magnetometer at zero applied magnetic field ($H_{\rm dc}$ = 0 T),  the signal of the FM anomaly was reduced below any detectable level at a pressure between 5.7 and 8.8 GPa~\cite{Gd_Tb_Dy_Ho_DAC_2021}. The anomaly at 5.7 GPa suggested a first order phase transition that remained in the range 0.5 - 11.6 GPa after remeasuring the sample by decreasing the pressure from 11.2 GPa (the results in the sequence after 11.2 GPa are not visible in Fig.~\ref{Figure-0} but details can be found in ref.~\cite{Gd_Tb_Dy_Ho_DAC_2021}).  Thus, residual strain influenced the suppression of the FM order.  

In our previous d.c.~magnetization measurements at $H_{\rm dc}$ = 0.5 T, which is one-third of the critical field to the saturated state, the FM anomaly was still observed at 7.3 GPa, and a broad hump appeared at 9.2 GPa~\cite{Gd_Tb_Dy_Ho_DAC}.   It suggests that the FM order becomes unstable above 7.3 GPa and a short range order composed of small grains is developed at 9.2 GPa. In different magnetization measurements, using a SQUID vibrating-coil-magnetometer method, the development of a ferromagnetic magnetization was observed at 8.2 GPa, which disappeared at 12 GPa~\cite{Gd_Tb_Dy_Ho_DAC_2021}.   All these facts together stressed  that the $P$ region of 8 - 12 GPa is a critical region for the FM ordering. 

Thus, different series of magnetic measurements showed that; i) the HM transition was observed at pressures up to 12 GPa, while the FM transition seems to be unstable at approximately 8 GPa, and ii) the ferromagnetic short range ordering could survive until approximately 11 GPa~\cite{Gd_Tb_Dy_Ho_DAC,Gd_Tb_Dy_Ho_DAC_2021}. 

However, previous neutron diffraction experiments, covering a $d$ spacing between 1.0 - 3.5 $\rm{\AA}$, reported that the FM ordering survives until at least 20 GPa~\cite{Ho_neutron_pressure}. Therefore, it seems that a controversy exists between the last neutron diffraction experiments by Perreault \textit{et al.}~\cite{Ho_neutron_pressure} and the macroscopic magnetic characterization of Ho at high pressures~\cite{Gd_Tb_Dy_Ho_DAC,Gd_Tb_Dy_Ho_DAC_2021}. 

To solve this controversy, in the present study, by performing neutron powder diffraction experiments at low temperatures, but covering a wide $d$-spacing region ($1.4\le d \le 50 \rm{\AA}$), we determined the magnetic structure at 8 GPa, in the temperature range $5\le T \le 300 \rm{K}$.  The magnetic superspace group (MSSG) formalism \cite{JANNER1980,PEREZMATO2012,RODRIGUEZCARVAJAL2019} has been employed to classify the symmetry of the magnetic structure.

\section{\label{sec:Exp} High Pressure Neutron Diffraction Experiments}

Polycrystalline sample of metallic natural Ho with high purity (99.999$\%$) was purchased from Sigma-Aldrich.  Especial care was taken to manipulate the sample minimizing the exposure time to air.

Neutron powder diffraction experiments were carried out on the high-flux 2-axis neutron diffractometer D1B of the Institut Laue-Langevin (ILL) in Grenoble, France. This instrument has a MWGC 1D-detector spanning an angular range of 128$^\circ$ with a definition of 0.1$^\circ$. A Radial Oscillating Collimator (ROC) was installed in order to eliminate the spurious signals produced by the sample environment.  

Two data acquisitions were taken at AP and room temperature (RT) for a Ho powder sample inside a 6 mm diameter vanadium can, with neutron wavelengths of $\lambda=1.28$  \AA\, and 2.52 \AA\, which allowed to explore $d$-spacings, respectively, of  0.7--15.0 \AA\  and 1.4--50 \AA.

The data collection at 8 GPa were performed with $\lambda=2.52$ \AA\ which correspond to the maximum flux configuration of the D1B instrument.   For these acquisitions, the powder was placed in a null-scattering TiZr gasket using a deuterated 4:1 ethanol-methanol mix as pressure transmitter medium, (the same that the one employed before by Perreault \textit{et al.} at 20 GPa~\cite{Ho_neutron_pressure}), which is the typical in all the neutron diffraction experiments. Then, it was introduced inside a VX5/180 Paris-Edinburgh (PE) pressure cell ~\cite{PE_cell_neutron,KLOTZ2005} equipped with SINE-type sintered diamond anvils \cite{KLOTZ2019}. A pressure of 0.12 GPa was applied to the PE cell, which for the sample corresponded to 8 GPa, after calibration with a Pb flake placed with the sample.  Then the PE cell was cooled using liquid nitrogen and helium from RT to 5 K. At this temperature, a 4.5 hours isotherm acquisition was performed. Then the sample was warmed back to RT in 10 hours and diffractograms were collected every 15 minutes.

It is well known that the 4:1 ethanol-methanol mixture is not the \textit{ideal} transmitting media at low temperatures but it is a good approach to the hydrostatic behavior at room temperature for pressures below 10 GPa \cite{otto_nonhydrostatic_1998}.  Furthermore, the fact that we do not detect any remarkable change, with decreasing the temperature, with the sample inside the PE cell, neither in the background of the diffractograms, nor in the width of the Bragg lines (see Fig.~\ref{8GPaC}), suggests that the quality of the pressure at 5 K is good enough in our experimental conditions.

Different crystallographic tools were employed for the determination of the crystal and magnetic structures, which include the FullProf Suite \cite{RODRIGUEZCARVAJAL1993}, the ISODISTORT Suite \cite{STOKES2017,CAMPBELL2006}, and utilities within the Bilbao Crystallographic Server \cite{AROYO2011,AROYO2006,AROYO20062,PEREZMATO2015} for the symmetry analysis and visualization. 
\section{\label{sec:Res}Results and Analysis}
\subsection{$P$ = 0 GPa}
\begin{figure}[htb!]
\begin{center}
\includegraphics[width=1\columnwidth]{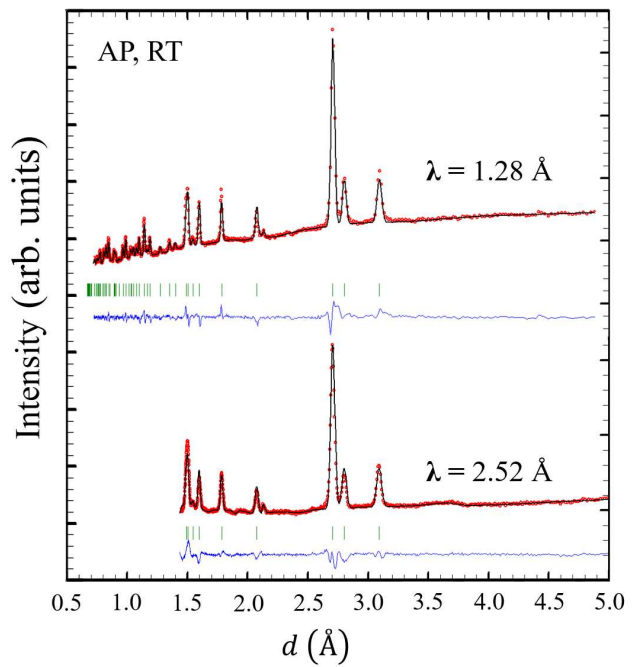}
\end{center}
\caption{\label{AP_RT} (Colour online) Diffraction patterns collected with $\lambda=1.28$~\AA\ (up) and 2.52 \AA\ (down) at RT and AP. In red are shown the observed points and in black continuous line the calculated pattern with the parameters described in the main text. The blue continuous line and the green ticks indicate, the difference between the observed and calculated profile, and the nuclear peaks generated by the space group, respectively.}
\end{figure}

Figure~\ref{AP_RT} shows the diffraction patterns collected with wavelengths $\lambda=1.28$ \AA\ (top) and 2.52 \AA\ (bottom).  The peaks observed in the diffraction patterns at AP and RT can be indexed by the paramagnetic $P6_3/mmc.1'$ MSSG (No. 194.264).  This crystal structure corresponds with a \textit{hcp} structure where the Ho atom is located in the Wyckoff position (WP) $2c$, with coordinates (1/3\ 2/3\ 1/4).   The cell parameters obtained from a multi-pattern refinement of both diffractograms are: $a = b = 3.5690(2)\ \rm{\AA},\,c = 5.6020(4)\,\rm{\AA},\,\alpha = \beta = 90^{\circ}, \gamma = 120^{\circ}$; with R$_{\rm Bragg}$ = 15.1 and 11.2, for $\lambda$ = 1.28 \AA\ and 2.52 \AA, respectively.
\subsection{$P$ = 8 GPa}
\begin{figure}[htb!]
\begin{center}
\includegraphics[width=1.\columnwidth]{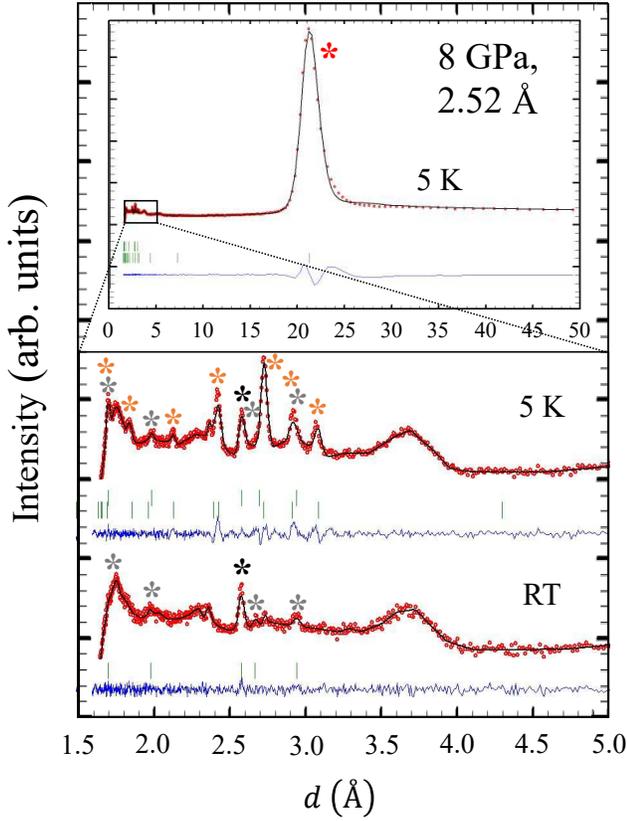}
\end{center}
\caption{\label{8GPaC} (Colour online). The insert shows the 5 K diffractogram in the full $d$-spacing scale accessible with the neutron wavelength $\lambda=2.52$ \AA\ at 8 GPa.  A large magnetic signal appears at $\sim 21$ \AA\,. A zoom of the small $d$-spacing region of the diffractograms collected with $\lambda=2.52$ \AA\, is displayed in the middle part (5K) and bottom part (RT) of the figure. In red are shown the observed points and in back continuous line the calculated pattern with the parameters described in the main text. The blue continuous line and the green ticks indicate the difference between the observed and calculated profile and the nuclear and satellite peaks generated by the magnetic superspace group, respectively. Red and orange asterisks label the satellite peaks visible only in the HM phase. Black and grey asterisks label the nuclear peaks visible at both temperatures. The broad feature around $3.7$ \AA\, and the small peak at $\sim 2.3$ \AA\, are due to the pressure transmitting fluid and the Pb flake, respectively, and treated as background contributions.}
\end{figure}

Insert of Fig.~\ref{8GPaC} shows the diffraction pattern collected with $\lambda=2.52$ \AA\ at $P= 8$ GPa and $T=5$ K. The first remarkable fact is the presence of a new high-intensity peak observed at $d\sim 21$ \AA\ that was not present in the diffractograms taken at RT.  The low part of the Fig.~\ref{8GPaC} shows a zoom of the small $d$-spacing region for the diffractograms collected at $T=5$ K (middle part) and RT (bottom part) with the sample inside the PE cell in the cryostat.  With this complex sample environment, once the sample is inside the gasket in the PE cell, the intensity is greatly suppressed, and a huge increase of the background is observed, even with the ROC in front of the detector. This is expected since for high pressure experiments, the quantity of sample irradiated by the neutron beam is much smaller. 

The diffractogram observed at RT inside the PE cell, after warming from 5K, is indexed by the same \textit{hcp} structure ($P6_3/mmc.1'$) with shorter cell parameters ($a=b=3.4030(7)$ \AA,\  $c= 5.345(2)$ \AA,\ and R$_{\rm Bragg}$ = 6.07) that the one observed at AP. Therefore, no structural transition to the Sm-type phase has been observed as the pressure was increased from AP to 8 GPa. 

However, at 8 GPa and 5 K, several new magnetic satellites appear, marked with red and orange asterisks in Fig.~\ref{8GPaC}, including the one clearly observed at $d\sim 21$~\AA.  These satellites are indexed with an incommensurate propagation vector $\vec{k}_{\rm{HM}}=(0\,0\,\delta)$ with $\delta=\pm 0.2536(1)$, in units of $c^*$, in agreement with Ref.~\cite{Ho_neutron_pressure}.  The plus or minus sign in the propagation vector indicates a clock-wise or anti clock-wise propagation.  The nuclear peaks, marked with black and grey asterisks in Fig.~\ref{8GPaC}, remain unchanged respect to the ones at RT, suggesting that no FM order is present, even at the lowest temperature.  The analysis of this new magnetic state will be elucidated in the next subsection.
\subsubsection{Symmetry analysis}
\begin{figure}[htb!]
\begin{center}
\includegraphics[width=1\columnwidth]{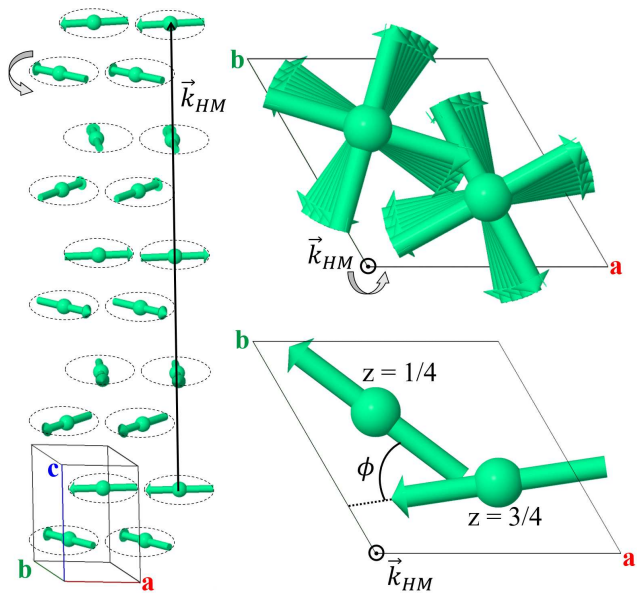}
\end{center}
\caption{\label{magstr} (Colour online) Magnetic structure obtained from the refinement of the data with the model given by the $P6_322.1'(0\,0\,\gamma)h00s$ MSSG. Five nuclear unit cells along $c$ axis are shown. The right bottom side shows the pitch angle $\phi$ formed by the magnetic moments in adjacent ferromagnetic $ab$ planes along $c$.  The right upper side shows a projection of the magnetic moments onto the $ab$ planes.}
\end{figure}

We use the symmetry analysis to reduce the number of possible magnetic ground states in Ho-\textit{hcp} compatible with the observed propagation vector $\vec{k}_{\rm{HM}}$. We can decompose the magnetic representation for the Ho atom, located at WP $2c$, as a direct sum of Irreducible Representations (\textit{irreps}) of the parent group $P6_3/mmc.1'$ for the $\Delta$ point, $(0\,0\,\delta)$, of the Brillouin zone (BZ) \footnote{We adopted the international notation for the \textit{irreps} labels and MSSG established in \cite{STOKES2017,CAMPBELL2006,AROYO2011}}, as follows:
\begin{eqnarray*}
 \rm{m}\Gamma_{2c}= 1m\Delta_2(1)\oplus 1m\Delta_3(1)\oplus 1m\Delta_5(2)\oplus 1m\Delta_6(2).
\label{eq:irreps}
\end{eqnarray*}
The basis vectors of each \textit{irrep} are given in Table \ref{tab:Irreps}. The magnetic structure described by both, the 1d-\textit{irreps} m$\Delta_2$ and m$\Delta_3$ consists of a sinusoidal modulation along the $c$-axis.  Meanwhile, the 2d-\textit{irreps}, m$\Delta_5$ and m$\Delta_6$, describe helices in which the magnetic moments are contained in ferromagnetic $ab$ planes, and propagate along the $c$-axis. However, both models differ greatly when considering the phase-shift between the Ho atoms in the unit cell. While in the case of m$\Delta_6$ the phase-shift is the same as the pitch angle of the helix: $\phi=180\times k_{\rm{HM}}=\pm45.65(2)^\circ$, in m$\Delta_5$ the phase-shift is given by: $\phi=180\times (k_{\rm{HM}}+1)=\pm225.65(2)^\circ$. Therefore, the magnetic structure given by m$\Delta_6$ can be considered as a single helix, while for m$\Delta_5$ the system is composed of two independent helices, one for each atom in the unit cell. The magnetic structure at 5 K, labelled with m$\Delta_6$, and the $\phi$ angle are depicted in Fig.~\ref{magstr}.

\begin{table}[h]
\begin{center}
\caption{Basis vectors (BV) of each \textit{irrep} present in the magnetic representation of the Ho $2c$ site for the space group $P6_3/mmc.1'$ with $\vec{k}_{\rm{HM}}=(0\,0\,\delta)$. The basis vectors components are expressed in terms of the crystallographic axis $a$, $b$ and $c$.\\}
\scalebox{0.95}{%
\begin{tabular}{|c|c|c|c|}
\hline 
\textit{irrep} & BV & Ho 1 (1/3\ 2/3\ 1/4) & Ho 2 (2/3\ 1/3\ 3/4) \\ \hline\hline
m$\Delta_2$ & $\psi _1^2$ & $\left(0 \,0\, 1\right)$ & $\left(0 \,0 \,1\right)\rm{e}^{-i\pi \delta}$ \\
m$\Delta_3$ & $\psi _1^3$ & $\left(0 \,0\, 1\right)$ & $\left(0 \,0 \,\bar{1}\right)\rm{e}^{-i\pi \delta}$ \\
m$\Delta_5$ & $\psi _1^5$ & $\left(1-\frac{1}{\sqrt{3}}\,i,\,-\frac{2}{\sqrt{3}}\,i,\,0\right)$ & $\left(1-\frac{1}{\sqrt{3}}\,i,\,-\frac{2}{\sqrt{3}}\,i,\,0\right) \rm{e}^{-i\pi (\delta+1)}$ \\
 & $\psi _2^5$ & $\left(-\frac{2}{\sqrt{3}}\,i,\,1-\frac{1}{\sqrt{3}}\,i,\,0\right)$ & $\left(-\frac{2}{\sqrt{3}}\,i,\,1-\frac{1}{\sqrt{3}}\,i,\,0\right) \rm{e}^{-i\pi (\delta+1)}$ \\
m$\Delta_6$ & $\psi _1^6$ & $\left(1-\frac{1}{\sqrt{3}}\,i,\,-\frac{2}{\sqrt{3}}\,i,\,0\right)$ & $\left(1-\frac{1}{\sqrt{3}}\,i,\,-\frac{2}{\sqrt{3}}\,i,\,0\right) \rm{e}^{-i\pi \delta}$ \\
 & $\psi _2^6$ & $\left(-\frac{2}{\sqrt{3}}\,i,\,1-\frac{1}{\sqrt{3}}\,i,\,0\right)$ & $\left(-\frac{2}{\sqrt{3}}\,i,\,1-\frac{1}{\sqrt{3}}\,i,\,0\right) \rm{e}^{-i\pi \delta}$ \\ \hline\hline
\end{tabular}
}
\label{tab:Irreps}
\end{center}
\end{table}

After a systematic trial and error procedure, it was observed that m$\Delta_2$, m$\Delta_3$ and m$\Delta_5$ do not fit the data, since they assign zero intensity for the main magnetic peak indexed as $(0\,0\,0)\pm\vec{k}_{\rm{HM}}$ at $d\sim 21$~\AA. However, the magnetic satellite reflections (red and orange asterisks in Fig.~\ref{8GPaC}) can be correctly fitted by the 2d-\textit{irrep} m$\Delta_6$.  

The combination between the parent group $P6_3/mmc.1'$ and the magnetic modulations for Ho atoms given by the \textit{irrep} m$\Delta_6$ give as a result the $P6_322.1'(0\,0\,\gamma)h00s$ MSSG. Within the MSSG formalism, the magnetic structure is described by a basic structure, related to the nuclear paramagnetic cell, in addition to a series of magnetic modulation functions that describe the variation from the basic structure of the magnetic moments. In our case, with just one propagation vector $\vec{k}_{\rm{HM}}$ and no net magnetic moment (i.e. no existence of a propagation vector $\vec{k}_{\rm{FM}}=(0\,0\,0)$), the magnetic structure is described by: 
\begin{eqnarray*}
\vec{M}_{lj} = \vec{M}_{j,\rm{s}}\sin\left( 2\pi x_4 \right) + \vec{M}_{j,\rm{c}}\cos\left( 2\pi x_4 \right),
\end{eqnarray*}
where the internal coordinate $x_4$ is given by the product of the propagation vector and the position $\vec{r}_{lj}$ of the atom $j$ in the $l$-th unit cell ($x_4 = \vec{k}_{\rm{HM}}\cdot\vec{r}_{lj}$). 

The explicit amplitudes of the cosine ($\vec{M}_{j,\rm{c}}$) and sine ($\vec{M}_{j,\rm{s}}$) components of the magnetic moment for the $P6_322.1'(0\,0\,\gamma)h00s$ MSSG are given in Table \ref{tab:MSSG}. In this MSSG, the Ho atom remains in the WP $2c$ position and the symmetry constraints imposed allow only for 1 free parameter to describe the magnetic structure, which is the modulus of the magnetic moment ($M$).

\begin{table}[h]
\begin{center}
\caption{Structural and magnetic parameters for Ho metallic at 8 GPa and $T=$300 and 5 K obtained from the analysis of the patterns collected at D1B. The cosine ($\vec{M}_{j,\rm{c}}$) and sine ($\vec{M}_{j,\rm{s}}$) components of the magnetic moment are expressed in terms of the crystallographic axis $a$, $b$ and $c$.\\}
\begin{tabular}{|c|cc|}
\hline
Phase & PM & HM \\
T (K) & 300 & 5 \\ \hline \hline 
MSSG & $P6_3/mmc.1'$ & $P6_322.1'(0\,0\,\gamma)h00s$ \\
\# & 194.264 & 182.1.24.2.m180.2 \\
$\vec{k}$ &  & $(0\,0\,\delta)$ \\
$\delta$  &  & 0.2536(1) \\
$\phi$ $(^\circ)$  &  & 45.65(2) \\
\textit{irrep}&  & m$\Delta_6$ \\ \hline \hline 
$a$ (\AA) & 3.4030(7) & 3.3976(6) \\
$c$ (\AA) & 5.345(2) & 5.390(1) \\ 
$c/a$ & 1.5707(7) & 1.5864(4) \\ \hline \hline 
$\vec{M}_{\rm{c}}$ &  & $M\left(0\,1\,0\right)$ \\
$\vec{M}_{\rm{s}}$ &  & $M\left(\frac{-2}{\sqrt{3}}\,\frac{-1}{\sqrt{3}}\,0\right)$\\ 
$M$($\mu_{\rm{B}}$) &  & 6.94(1) \\ \hline \hline 
R$_{\rm{Bragg}}$ & 6.07 & 6.64 \\ 
R$_{\rm{Bragg}}$(Mag.) &  & 1.59 \\   \hline \hline
\end{tabular}
\label{tab:MSSG}
\end{center}
\end{table}

\subsubsection{Temperature dependence}
\begin{figure}[htb!]
\begin{center}
\includegraphics[width=0.95\columnwidth]{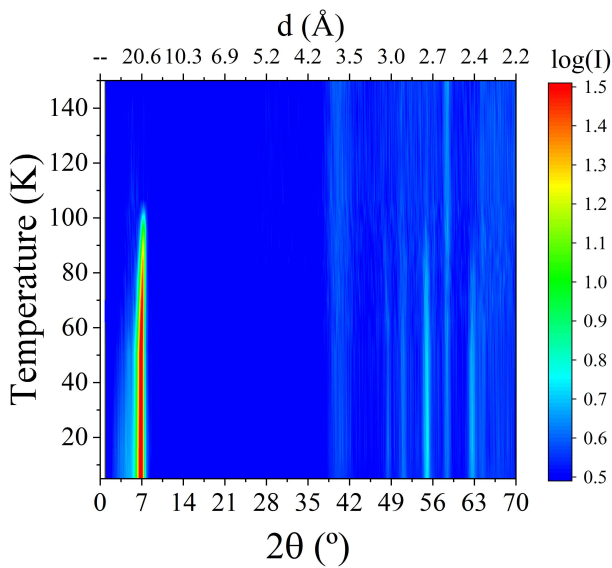}
\end{center}
\caption{\label{2Dplot} (Colour online) 2D plot of the thermo-diffractograms measured at 8 GPa with $\lambda =2.52$ \AA\, as the system was heated from 5 K to RT. The appearance of the peak at $d\sim 21 \rm{\AA}$ ($2\theta\sim 7^\circ$) is clearly visible at $T=$ 105(2) K.}
\end{figure}

Figure~\ref{2Dplot} shows a 2D plot of the thermo-diffractograms obtained at 8 GPa as the system was heated from 5 K to RT. From them, the onset of the helical magnetic ordering is estimated to occur around $T_{\rm N}$ = 105(2) K,  with the appearance of a peak at $d\sim 21 \rm{\AA}$ ($2\theta\sim 7^\circ$), which is in agreement with previous studies \cite{Gd_Tb_Dy_Ho_DAC_2021, Jackson_2005, Gd_Tb_Dy_Ho_DAC, Ho_neutron_pressure}.

\begin{figure}[htb!]
\begin{center}
\includegraphics[width=1\columnwidth]{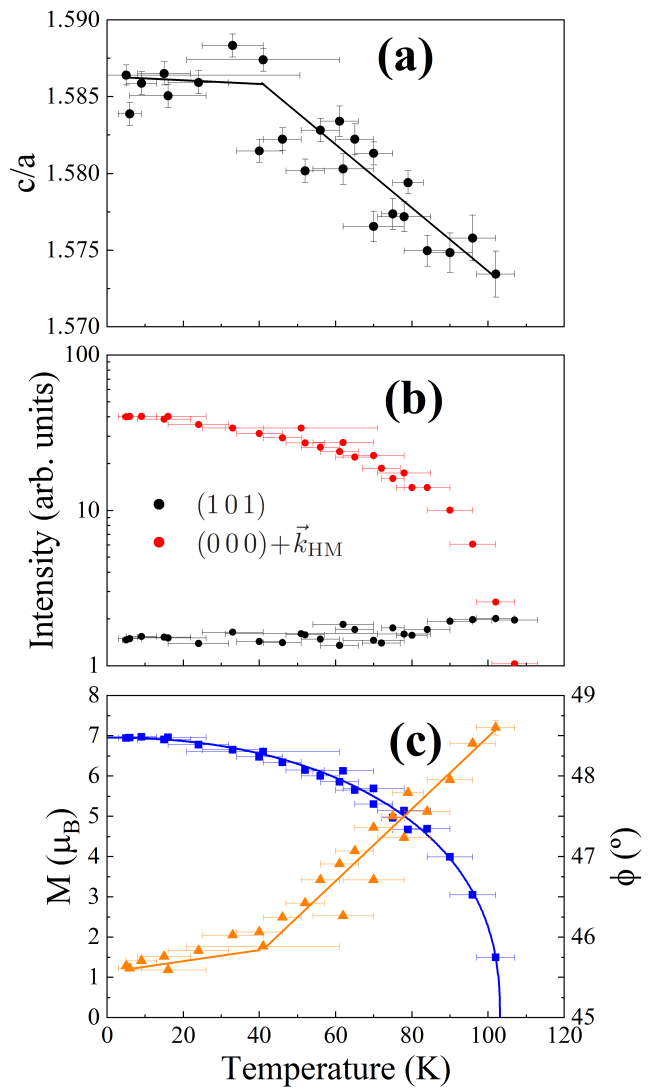}
\end{center}
\caption{\label{Tevolution} (Colour online) (a) Temperature dependence of the ratio $c/a$. (b) Logarithmic scale plot of the intensity evolution with temperature for the nuclear peak $(1\,0\,1)$ (black) and the satellite reflection $(0\,0\,0)\pm\vec{k}_{\rm{HM}}$ (red). These peaks are marked with black and red asterisks in the diffractograms  of Fig.~\ref{8GPaC}. (c) Temperature dependence of the modulus $M$($\mu_{\rm B}$) (blue squares) and $\phi$ angle values (orange triangles). The lines are a guide to the eyes.  A change in the linear behaviour of $\phi$ is observed at around 40 K.}
\end{figure}

The temperature dependence of the intensity for the (1\,0\,1) nuclear reflection (in black) and the satellite $(0\,0\,0)\pm\vec{k}_{\rm{HM}}$ (in red) is shown in Fig.~\ref{Tevolution}(b). The evolution of these two lines allows to quantitatively distinguish between the magnetic phases that can be present, as the nuclear and satellite peaks chosen should be the most sensible if a FM ($\vec{k}_{\rm{FM}}=(0\,0\,0)$) or HM order ($\vec{k}_{\rm{HM}}$), respectively, should be present. No change is observed neither in the (1\,0\,1) nuclear reflection, nor other nuclear lines, as the temperature decreases. This fact supports the hypothesis of absence of any FM ordering, or if present, put an upper limit of 0.2$\mu_{\rm B}$ to such contribution.

\begin{figure}[htb!]
\begin{center}
\includegraphics[width=0.95\columnwidth]{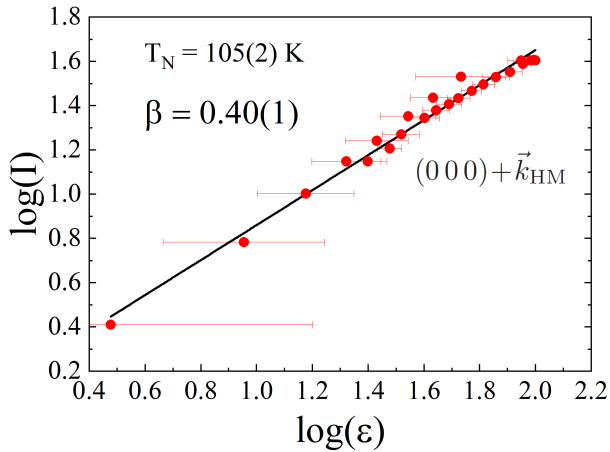}
\end{center}
\caption{\label{Beta} (Colour online) Logarithmic plot of the intensity of the satellite reflection $(0\,0\,0)\pm\vec{k}_{\rm{HM}}$ as a function of the reduced temperature. The value of the transition temperature $T_{\rm N}$ was estimated from Fig.~\ref{2Dplot}.}
\end{figure}

Regarding the evolution with temperature of the satellite $(0\,0\,0)\pm\vec{k}_{\rm{HM}}$, its intensity was fitted to the power law $I \propto \epsilon^{2\beta}$, where $\epsilon = T_{\rm N}-T$ is the reduced temperature (see Fig.~\ref{Beta}). The obtained value for the critical exponent is $\beta=0.40(1)$, which is in good agreement with the value predicted by Bak and Mukamel \cite{BAK1976}, and reported by previous studies for pure helimagnetic phases at AP and 130 K conditions \cite{ECKERT1976,THURSTON1993,THURSTON1994,PLATHKY2001}.  It constitutes another indication of the absence of any FM component.

Using the magnetic model labelled by the $P6_322.1'(0\,0\,\gamma)h00s$ MSSG to fit the diffraction patterns collected at each temperature, the dependence of the pitch angle, $\phi$, and modulus, $M$, of the Ho magnetic moment versus temperature was obtained (see Fig.~\ref{Tevolution}(c)). 

As the temperature decreases below $T_{\rm N}\sim105(2)$~K, the magnetic moment increases until it saturates at 6.94(1)$\mu_{\rm B}$, a value in good agreement with the saturation magnetization obtained at AP and $T = 5$ K \cite{Gd_Tb_Dy_Ho_DAC}. Meanwhile, the $\phi$ angle decreases following two linear dependences from 48.6(1)$^\circ$ at 105 K to 45.65(2)$^\circ$ at 5K, with a change of slope around $T\sim 40$ K. This evolution is also coherent with previous neutron diffraction experiments \cite{Ho_neutron_pressure}, and can be understood if we consider the strong dependence of the helix period with the axial ratio $c/a$ of the hexagonal phase \cite{HUGHES2007}. Such dependence can also explain the change observed around 40 K, since the ratio $c/a$ is stabilized below that temperature (see Fig.~\ref{Tevolution}(a)).

\section{\label{sec:Discussion}Discussion}
In our last study using SQUID magnetization measurements it was reported that the HM order survives up to $P=12$ GPa, while the disappearance of the FM order, or, at least, a remarkable suppression of ferromagnetic net magnetic moments, was observed above 8 GPa \cite{Gd_Tb_Dy_Ho_DAC_2021}. As the intensity of the FM anomaly could not be detected just above the critical pressure for the phase boundary between the \textit{hcp} and Sm-type phases, it was suggested that the disappearance of the FM order is related to the structural phase transition.

However, the present neutron diffraction experiments, focused on the magnetic structure at $P=8$ GPa, confirm that the \textit{hcp} structure remains stable until at least 8 GPa. Additionally, an incommensurate HM order is observed below $T_{\rm N}=105(2)$ K, which persists down to the lowest temperature measured ($T=5$ K), while no evidence of FM ordering was observed within the entire temperature range. Therefore, the present results help to clarify that the FM order becomes unstable and disappear just before the structural phase transition. Furthermore, the magnetic structure observed at 8 GPa is consistent with previous magnetic measurements, taking into account the pressure distribution with the order of $\pm$0.5 GPa during the high-pressure experiment.

In the neutron diffraction study by Perreault \textit{et al.}, the FM order was reported to survive in both the Sm-type phase at 14.2 GPa and the \textit{dhcp} phase at 20.2 GPa \cite{Ho_neutron_pressure}. This FM transition was marked by the appearance of a magnetic peak at $d=3$ \AA\ accompanied by an increase in the intensity of all nuclear peaks. However, in those experiments only a small region of $d$-spacing (1.0--3.5 \AA) was covered, which made it difficult to elucidate the additional existence of HM order at those pressures, since the expected most intense magnetic signal should appear at larger $d$-spacing (see Fig. \ref{8GPaC}). Therefore, it would be interesting to confirm these results by performing neutron experiments at such pressures in a long $d$-spacing diffractometer.
\section{\label{sec:Conclusion}Conclusion}
We conducted neutron diffraction experiments to investigate the magnetic structures of Ho at $P= 8$ GPa. Our findings indicate that the nuclear symmetry remains unchanged, preserving the hexagonal close-packed ($hcp$) symmetry at $P=8$ GPa. The helimagnetic order persists down to 5 K, and the analysis of its magnetic structure, using the magnetic superspace group formalism, allows to determine the critical exponent $\beta=0.40(1)$, in agreement with previous theoretical model for pure helices. Meanwhile no FM contribution is observed at any temperature. These results are consistent with previously published findings from magnetization experiments.
\\
\begin{acknowledgments}
This work was supported by Grants-in-Aid for Scientific Research, Grant No. 19KK0070, from the Ministry of Education, Culture, Sports, Science and Technology (MEXT), Japan.   The authors acknowledge support by Grants No. PID2022-138492NB-I00-XM4 funded by MCIN/AEI/10.13039.501100011033, OTR02223-SpINS from CSIC/MCIN and DGA/M4 from Diputaci\'on General de Arag\'on (Spain). MPS acknowledges a predoctoral research fellowship from Diputaci\'on General de Arag\'on (Spain). Authors acknowledge the SANE service of ILL and in particular to C. Payre for his valuable support during the experiments. 
\end{acknowledgments}
%
\bibliography{Ref_Mito}
%
\end{document}